\documentclass[twocolumn,english,american,pra,aps,showpacs]{revtex4}
\usepackage{graphicx}
\usepackage{amssymb}

\makeatletter

\usepackage{graphicx}

\usepackage{bm}

\usepackage{babel}
\makeatother
\begin{document}

\title{Conclusive and arbitrarily perfect quantum state transfer using parallel
spin chain channels}

\author{Daniel Burgarth and Sougato Bose}

\affiliation{Department of Physics \& Astronomy, University College London, Gower
St., London WC1E 6BT, UK}

\begin{abstract}
We suggest a protocol for perfect quantum communication through spin
chain channels. By combining a dual-rail encoding with measurements
only at the receiving end, we can get conclusively perfect state transfer,
whose probability of success can be made arbitrarily close to unity.
As an example of such an \emph{amplitude delaying channel}, we show
how two parallel Heisenberg spin chains can be used as quantum wires.
Perfect state transfer with a probability of failure lower than $P$
in a Heisenberg chain of $N$ spin-$1/2$ particles can be achieved
in a timescale of the order of $\frac{0.33\hbar}{J}N^{1.7}|\ln P|$.
We demonstrate that our scheme is more robust to decoherence and non-optimal
timing than any scheme using single spin chains. 
\end{abstract}


\maketitle

\section{Introduction}

The development of reliable methods to transfer quantum states is
of fundamental importance in quantum information theory. Usually,
flying qubits such as photons, ballistic electrons and guided atoms/ions
are considered for this purpose. However, converting back and forth
between stationary qubits (of say a quantum computer or as held by
the communicating parties) and mobile carriers of quantum information
and interfacing between different physical implementations of qubits
is very difficult and may not be really worth for short communication
distances.

An attractive alternative is to use a finite array of interacting
but stationary qubits as an information bus. It would, however, not
be very useful if the interactions between various pairs of stationary
qubits have to be repeatedly switched on and off to perform the communication,
because gating errors will then accumulate. Moreover, the local control
required will be as high as that of a quantum computer. Only if we
can utilize systems with permanently coupled material qubits (such
as molecular spin chains), or systems without local control (such
as Josephson junction arrays or optical lattices) with minimal global
switchings, can we have a communication bus much before a quantum
computer. In such schemes, both the amplitude and the phase damping
(mechanisms of decoherence) can be ensured to be no worse than that
of a single moving qubit, by ensuring that there is at most one excitation
in the array during the communication process. With the above view
in mind, recently the use of spin chains \cite{SB03,key-1,MCND+04,key-2,TJO04,key-3,FVMAM04,key-4,key-5,MYDL04,key-6,key-7}
and harmonic chains \cite{PLENIO04} as quantum wires have been proposed.

The initial system independent proposal \cite{SB03} was inspired
by the natural setting for spin chain molecules (and optical lattices):
regular arrays without local accessibility. Single spin encoding was
assumed to avoid quantum gates. For this simplicity, its specific
realizations is already being proposed \cite{JOSEPH}. However, it
allows only imperfect communication fidelity and necessitate the use
of entanglement distillation from a large ensemble, which destroy
its simplicity. Later approaches of perfecting fidelity require either
the engineering of the couplings \cite{MCND+04,key-2}, or an encoding
of a qubit to several spins \cite{TJO04,key-3}. Independently, local
measurements on each qubit along the chain \cite{FVMAM04,key-4,key-5}
have been proposed. The above deviate either from naturalness or from
simplicity. We are thus sorely in need of a scheme which remains natural
and simple, yet achieves perfect quantum communication. This is achieved
in this paper by using a dual-rail encoding.

The outline of the article is as follows. In Section \ref{sec:Scheme-for-conclusive},
we suggest a scheme for quantum communication using two parallel spin
chains of the most natural type (namely those with constant couplings).
We require modest encodings (or gates) and measurements only at the
ends of the chains. The state transfer is \emph{conclusive}, which
means that it is possible to tell by the outcome of a quantum measurement,
without destroying the state, if the transfer took place or not. If
it did, then the transfer was \emph{perfect}. The transmission time
for conclusive transfer is no longer than for single spin chains.
In Section \ref{sec:Arbitrarily-perfect-state}, we demonstrate that
our scheme offers even more: if the transfer was not successful, then
we can wait for some time and just repeat the measurement, without
having to resend the state. By performing sufficiently many measurements,
the probability for perfect transfer approaches unity. Hence the transfer
is \emph{arbitrarily perfect}. We will show in Section \ref{sec:Estimation-of-the}
that the time needed to transfer a state with a given probability
scales in a reasonable way with the length of the chain. Finally,
in Section \ref{sec:Decoherence-and-imperfections} we show that encoding
to parallel chains and the conclusiveness also makes our protocol
more robust to decoherence (a hitherto unaddressed issue in the field
of quantum communication through spin chains).%
\begin{figure*}
\begin{center}\includegraphics[%
  width=0.70\textwidth]{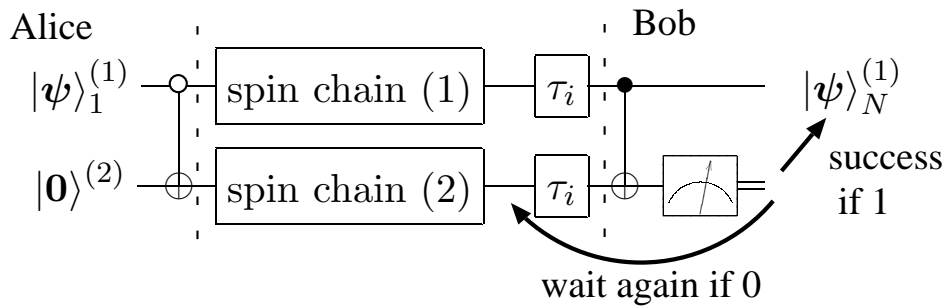}\end{center}

\caption{\label{fig:summary}Quantum circuit representation of conclusive
and arbitrarily perfect state transfer. The first gate at Alice's
qubits represents a NOT gate applied to the second qubit controlled
by the first qubit being zero. The qubits $\left|\bm{\psi}\right\rangle _{1}^{(1)}$
on the left hand side represents an arbitrary input state at Alice's
site, and the qubit $\left|\bm{\psi}\right\rangle _{N}^{(1)}$ represents
the same state, successfully transferred to Bob's site. The $\tau_{i}$-gate
represents the unitary evolution of the spin chains for a time interval
of $\tau_{i}$.}
\end{figure*}

\section{Scheme for conclusive transfer\label{sec:Scheme-for-conclusive}}

We intend to propose our scheme in a system independent way with occasional
references to systems where conditions required by our scheme are
achieved. We assume that our system consists of two identical uncoupled
spin-$1/2$-chains $(1)$ and $(2)$ of length $N$, described by
the Hamiltonian\begin{equation}
H=H^{(1)}\otimes I^{(2)}+I^{(1)}\otimes H^{(2)}-E_{g}I^{(1)}\otimes I^{(2)}.\end{equation}
 The term identical states that $H^{(1)}$ and $H^{(2)}$ are the
same apart from the label of the Hilbert space they act on. The requirement
of parallel chains instead of just one is not problematic, since in
many experimental realizations of spin chains, it is much easier to
produce a whole bunch of parallel uncoupled \cite{NMEE96,GAMBARDELLA}
chains than just a single one. 

We assume that the ground state of each chain is $\left|\mathbf{0}\right\rangle ^{(i)}\equiv\left|0_{1}\ldots0_{N}\right\rangle ^{(i)}$,
i.e. a ferromagnetic ground state, with $H^{(i)}\left|\mathbf{0}\right\rangle ^{(i)}=E_{g}\left|\mathbf{0}\right\rangle ^{(i)},$
and that the subspace consisting of the single spin excitations $\left|\mathbf{n}\right\rangle ^{(i)}\equiv\sigma_{n}^{+(i)}\left|\mathbf{0}\right\rangle ^{(i)}\,(n=1,\ldots,N)$
is invariant under $H^{(i)}.$ An arbitrary qubit at the site $n$
of system $(i)$ can be written as\begin{equation}
\left|\bm{\psi}\right\rangle _{n}^{(i)}\equiv\alpha\left|\mathbf{0}\right\rangle ^{(i)}+\beta\left|\mathbf{n}\right\rangle ^{(i)}.\end{equation}
 The dynamics restricted to this subspace can be expressed in terms
of the transition amplitudes\begin{equation}
f_{r,s}(t)\equiv\left\langle \mathbf{r}\right|^{(i)}e^{iH^{(i)}t}\left|\mathbf{s}\right\rangle ^{(i)}.\end{equation}
 The aim of our protocol is to transfer quantum information from the
$1$st ({}``Alice'') to the $N$th ({}``Bob'') qubit of the first
chain,\begin{equation}
\left|\bm{\psi}\right\rangle _{1}^{(1)}\longrightarrow\left|\bm{\psi}\right\rangle _{N}^{(1)}.\end{equation}
 To achieve this, we need $f_{1,N}(t)\neq0$ (which is valid for a
Heisenberg chain, for example \cite{SB03}). An advantage of Heisenberg
ferromagnetic chains over a non-interacting qubit array is that some
XXZ anisotropy can make the states $\left|\mathbf{0}\right\rangle $
and $\left|\mathbf{n}\right\rangle $ stable against excitations at
finite temperatures \cite{key-8}. Even a small anisotropy in the
coupling $J$ may suffice (as $J$ itself can be as high as $2000K$
\cite{NMEE96}). Alternatively, one can prevent thermal excitations
by applying an uniform magnetic field to the chain.

The initial state of the system is $\left|\bm{\psi}\right\rangle _{1}^{(1)}\otimes\left|\mathbf{0}\right\rangle ^{(2)}$.
The first step of the protocol is to encode the input qubit in a {}``dual-rail''
\cite{ILC96} by applying a NOT gate on the first qubit of system
$(2)$ controlled by the first qubit of system $(1)$ being zero,
resulting in a superposition of excitations in both systems, \begin{equation}
\left|\mathbf{\bm{s}}(0)\right\rangle =\alpha\left|\mathbf{0}\right\rangle ^{(1)}\otimes\left|\mathbf{1}\right\rangle ^{(2)}+\beta\left|\mathbf{1}\right\rangle ^{(1)}\otimes\left|\mathbf{0}\right\rangle ^{(2)}.\label{eq:superposition}\end{equation}
 This is assumed to take place in a much shorter timescale than the
system dynamics. Even though a 2-qubit gate in solid state systems
is difficult, such a gate for charge qubits has been reported \cite{TYYAP+03}.
For the same qubits, Josephson arrays have been proposed as single
spin chains for quantum communication \cite{JOSEPH}. For this system,
both requisites of our scheme are thus available. In fact, the demand
that Alice and Bob can do measurements and apply gates to their local
qubits (i.e. the ends of the chains) will be naturally fulfilled in
practice since we are suggesting a scheme to transfer information
between quantum computers.

Under the system Hamiltonian, the excitation in Eq. (\ref{eq:superposition})
will travel along the two systems. The state after the time $\tau_{1}$
can be written as \begin{equation}
\left|\mathbf{\bm{\phi}}(\tau_{1})\right\rangle =\sum_{n=1}^{N}f_{n,1}(\tau_{1})\left|\mathbf{s}(n)\right\rangle ,\end{equation}
 where $\left|\mathbf{s}(n)\right\rangle =\alpha\left|\mathbf{0}\right\rangle ^{(1)}\otimes\left|\mathbf{n}\right\rangle ^{(2)}+\beta\left|\mathbf{n}\right\rangle ^{(1)}\otimes\left|\mathbf{0}\right\rangle ^{(2)}$.
We can decode the qubit by applying a CNOT gate at Bob's site. The
state thereafter will be \begin{equation}
\sum_{n=1}^{N-1}f_{n,1}(\tau_{1})\left|\mathbf{s}(n)\right\rangle +f{}_{N,1}(\tau_{1})\left|\bm{\psi}\right\rangle _{N}^{(1)}\otimes\left|\mathbf{N}\right\rangle ^{(2)}.\end{equation}
 Bob can now perform a measurement on his qubit of system $(2).$
If the outcome of this measurement is $1$, he can conclude that the
state $\left|\bm{\psi}\right\rangle _{1}^{(1)}$ has been successfully
transferred to him. This happens with the probability $\left|f_{N,1}(\tau_{1})\right|^{2}.$
If the outcome is $0$, the system is in the state \begin{equation}
\frac{1}{\sqrt{P(1)}}\sum_{n=1}^{N-1}f_{n,1}(\tau_{1})\left|\mathbf{s}(n)\right\rangle ,\label{eq:firstmeas}\end{equation}
where \begin{equation}
P(1)=1-\left|f_{N,1}(\tau_{1})\right|^{2}\end{equation}
is the probability of failure for the first measurement. If the protocol
stopped here, and Bob would just assume his state as the transferred
one, the channel could be described as an amplitude damping channel,
with exactly the same fidelity as the single chain scheme discussed
in \cite{SB03}. But success probability is more valuable than fidelity:
Bob has gained knowledge about his state, and may reject it and ask
Alice to retransmit. However, as we will show in the next section,
this is not necessary.

\section{Arbitrarily perfect state transfer\label{sec:Arbitrarily-perfect-state}}

\begin{figure}
\begin{center}\includegraphics[%
  width=1.0\columnwidth]{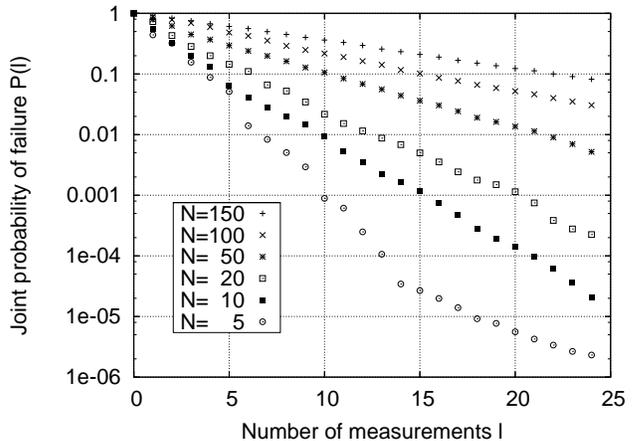}\end{center}

\caption{\label{fig:heisenberg}Semilogarithmic plot of the joint probability
of failure $P(l)$ as a function of the number of measurements $l$.
Shown are Heisenberg spin-$1/2$-chains with different lengths $N$.
The times between measurements $\tau_{i}$ have been optimized numerically. }
\end{figure}

Because Bob's measurement has not revealed anything about the input
state, the information is still residing in the chain. By letting
the state (\ref{eq:firstmeas}) evolve for another time $\tau_{2}$
and applying the CNOT gate again, Bob has another chance of receiving
the input state. The state before performing the second measurement
is easily seen to be \begin{equation}
\frac{1}{\sqrt{P(1)}}\sum_{n=1}^{N}\left\{ f_{n,1}(\tau_{2}+\tau_{1})-f_{n,N}(\tau_{2})f_{N,1}(\tau_{1})\right\} \left|\mathbf{s}(n)\right\rangle .\label{eq:second_meas}\end{equation}
 Hence the probability to receive the qubit at Bobs site at the second
measurement is\begin{equation}
\frac{1}{P(1)}\left|f_{N,1}(\tau_{2}+\tau_{1})-f_{N,N}(\tau_{2})f_{N,1}(\tau_{1})\right|^{2}.\end{equation}
 If the transfer was still unsuccessful, this strategy can be repeated
over and over. Each time Bob has a probability of failed state transfer
that can be obtained from the generalization of Eq. (\ref{eq:second_meas})
to an arbitrary number of iterations. The joint probability that Bob
fails to receive the state all the time is just the product of these
probabilities. We denote the joint probability of failure for having
done $l$ unsuccessful measurements as $P(l)$. This probability depends
on the time intervals $\tau_{i}$ between the $(i-1)$th and $i$th
measurement, and we are interested in the case where the $\tau_{i}$
are chosen such that the transfer is fast. It is possible to write
a simple algorithm that computes $P(l)$ for any transition amplitude
$f_{r,s}(t).$ Figure \ref{fig:heisenberg} shows some results for
a Heisenberg spin-$1/2$-chain with equal nearest neighbor couplings,
\begin{equation}
H^{(i)}=-J\sum\vec{\sigma}_{n}^{(i)}\cdot\vec{\sigma}_{n+1}^{(i)}.\end{equation}
 This model is exactly solvable, and the transition amplitude is given
explicitly in \cite{SB03}. However, the results are valid for a wide
class of anisotropies and in the presence of a uniform magnetic field,
too.%

An interesting question is whether the joint probability of failure
can be made arbitrary small with a large number of measurements. Since
$P(l)$ is a bounded, monotonic decreasing series, it must have a
limit. In fact, the times $\tau_{i}$ can be chosen such that the
transfer becomes arbitrarily perfect. This has been proven in \cite{BBG},
where a generalization of the above scheme and a much wider class
of Hamiltonians is considered. In the limit of large number of measurements,
the spin channel will not damp the initial amplitude, but only \emph{delay}
it.

\section{Estimation of the timescale the transfer\label{sec:Estimation-of-the}}

The achievable fidelity is an important, but not the only criteria
of a state transfer protocol. In this Section, we give an heuristic
approach to estimate the time that it needs to achieve a certain fidelity
in a Heisenberg spin chain. The comparison with numeric examples is
confirming this approach. 

\begin{figure}
\includegraphics[%
  width=1.0\columnwidth]{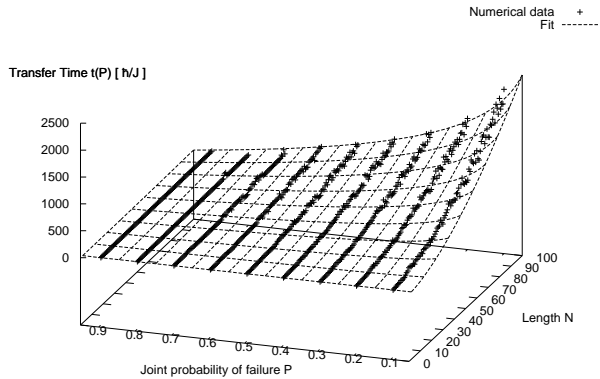}

\caption{\label{cap:Timefit}Time $t$ needed to transfer a state with a given
joint probability of failure $P$ across a chain of length $N$. The
points denote exact numerical data, and the fit is given by Eq. (\ref{eq:fit}).}
\end{figure}

Let us first describe the dynamic of the chain in a very qualitative
way. Once Alice has initialized the system, an excitation wave packet
will travel along the chain. As shown in \cite{SB03}, it will reach
Bob at a time of the order of \begin{equation}
\tau_{max}\approx\frac{\hbar N}{2J},\end{equation}
 with an amplitude of \begin{equation}
\left|f_{N,1}(\tau_{max})\right|^{2}\approx1.35N^{-2/3}.\label{eq:peak}\end{equation}
It is then reflected and travels back and forth along the chain. Since
the wave packet is also dispersing, it starts interfering with its
tail, and after a couple of reflections the dynamic is becoming quite
randomly. This effect becomes even stronger due to Bobs measurements,
which change the dynamics by projecting away parts of the wave packet.
However, $2\tau_{max}$ (the time it takes for a wave packet to travel
twice along the chain) remains a good estimate of the timescale in
which significant probability amplitude peaks at Bobs site occur,
and Eq. (\ref{eq:peak}) remains a good estimate of the amplitude
of these peaks. Therefore, the joint probability of failure is expected
to scale as\begin{equation}
P(l)\approx\left(1-1.35N^{-2/3}\right)^{l}\label{eq:probscale}\end{equation}
in a time of the order of \begin{equation}
t(l)\approx2\tau_{max}l=\frac{N\hbar l}{J}.\label{eq:timescale}\end{equation}
If we combine Eq. (\ref{eq:probscale}) and (\ref{eq:timescale})
and solve for the time $t(P)$ needed to reach a certain probability
of failure $P$, we get \begin{equation}
t(P)\approx\frac{0.51\hbar N^{5/3}}{J}\left|\ln P\right|.\end{equation}
We compare this rough estimate with exact numerical results in Fig.
\ref{cap:Timefit}%
. The best fit is given by\begin{equation}
t(P)=\frac{0.33\hbar N^{5/3}}{J}\left|\ln P\right|.\label{eq:fit}\end{equation}
\foreignlanguage{english}{We can conclude that the transmission time
for arbitrarily perfect transfer is scaling not much worse with the
length $N$ of the chains than the single spin chain schemes. Despite
of the logarithmic dependence on $P,$ the time it takes to achieve
high fidelity is still reasonable. For example, a system with $N=100$
and $J=20K*k_{B}$ will take approximately $1.3ns$ to achieve a fidelity
of 99\%. In many systems, decoherence is completely negligible within
this timescale. For example, some Josephson junction systems \cite{key-11}
have a decoherence time of $T_{\phi}\approx500ns$, while trapped
ions have even larger decoherence times.}

\section{Decoherence and imperfections\label{sec:Decoherence-and-imperfections}}

If the coupling between the spins $J$ is very small, or the chains
are very long, the transmission time may no longer be negligible with
respect to the decoherence time (see Section \ref{sec:Estimation-of-the}).
It is interesting to note that the dual-rail encoding then offers
some significant general advantages over single chain schemes. Since
we are suggesting a system-independent scheme, we will not study the
effects of specific environments on our protocol, but just qualitatively
point out its general advantages.

At least theoretically, it is always possible to cool the system down
or to apply a strong magnetic field such that the environment is not
causing further excitations. Then, there are two remaining types of
quantum noise that will occur: phase noise and amplitude damping.
Phase noise is a serious problem and arises here \emph{only} when
an environment that can distinguish between spin flips on the first
chain and spin flips on the second chain. It is therefore important
that the environment cannot resolve their difference. In this case,
the environment will only couple with the total $z$-component\begin{equation}
S_{z,n}=\sigma_{z,n}^{(1)}+\sigma_{z,n}^{(2)}\end{equation}
 of the spins of both chains at each position $n$. This has been
discussed for spin-boson models in \cite{key-9,key-10} but should
also hold for spin environments as long as the chains are close enough.
The qubit is encoded in a decoherence-free subspace \cite{DFSUB}
and the scheme is fully robust to phase noise. Even though this may
not be true for all implementations of dual-rail encoding, it is worthwhile
noticing it because such an opportunity does not exist \emph{at all}
for single chain schemes, where the coherence between two states with
different total z-component of the spin has to be preserved. Having
shown one way of avoiding phase noise, at least in some systems, we
now proceed to amplitude damping.

\begin{figure}
\begin{center}\includegraphics[%
  width=1.0\columnwidth]{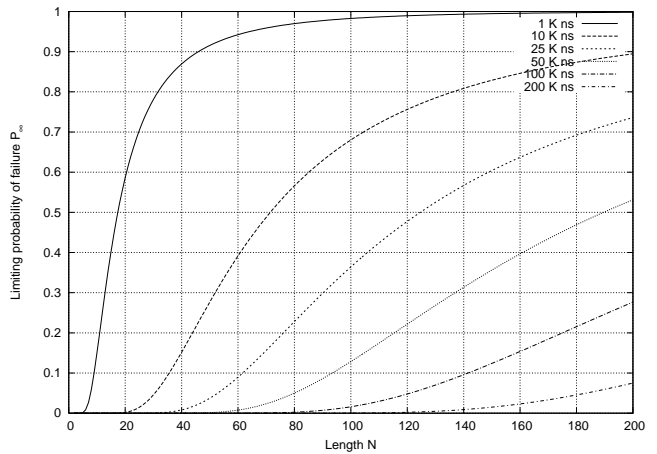}\end{center}

\caption{\label{fig:losses}The minimal joint probability of failure $P(l)$
for chains with length $N$ in the presence of amplitude damping.
The parameter $J/\Gamma$ of the curves is the coupling of the chain
(in Kelvin) divided by the decay rate ($ns^{-1}$).}
\end{figure}

The evolution of the system in presence of amplitude damping of a
rate $\Gamma$ can be easily derived using a quantum-jump approach
\cite{QJUMP}. Like for phase noise, it is necessary that the environment
acts symmetrically on the chains. The dynamics is then given by an
effective non-unitary Hamiltonian\begin{equation}
H_{eff}=H+i\Gamma\sum_{n}\left(S_{z,n}+2\right)/2\end{equation}
if no jump occurs, and the effect of a jump is given by the operator\begin{equation}
\sum_{n}S_{n}^{-},\end{equation}
which will put the system in the ground state. As this can be solved
analytically, we do not go into numerics. The state of the system
before the first measurement conditioned on no jump is given by\begin{equation}
e^{-\Gamma t}\sum_{n=1}^{N}f_{n,1}(t)\left|\mathbf{s}(n)\right\rangle ,\label{eq:jump1}\end{equation}
and this happens with the probability of $e^{-2\Gamma t}$ (the norm
of the above state). If a jump occurs, the system will be in the ground
state \begin{equation}
\sqrt{1-e^{-2\Gamma t}}\left|\mathbf{0}\right\rangle ^{(1)}\otimes\left|\mathbf{0}\right\rangle ^{(2)}.\label{eq:jump2}\end{equation}
The density matrix at the time $t$ is given by a mixture of (\ref{eq:jump1})
and (\ref{eq:jump2}). In case of (\ref{eq:jump2}), the quantum information
is completely lost and Bob's error check qubit will never show success.
If Bob however measures a success, it is clear that no jump has occurred
and he has the perfectly transferred state. Therefore the protocol
\emph{remains conclusive}, but the success probability is lowered
by $e^{-2\Gamma t}.$ This result is still valid for multiple measurements,
which leave the state (\ref{eq:jump2}) unaltered. The probability
of a successful transfer at each particular measurement $l$ will
decrease by $e^{-2\Gamma t(l)}$, where $t(l)$ is the time of the
measurement. After a certain number of measurements, the \emph{joint}
probability of failure will no longer decrease. Thus the transfer
will no longer be \emph{arbitrarily} perfect, but can still reach
a very high fidelity. Some numerical examples of the minimal joint
probability of failure that can be achieved, \begin{eqnarray}
P_{\infty} & \equiv & \lim_{l\rightarrow\infty}P(l)\\
 & \approx & \prod_{l=1}^{\infty}\left(1-1.35N^{-2/3}e^{-\frac{2\Gamma N\hbar}{J}l}\right)\label{eq:plimit}\end{eqnarray}
are given in Fig. \ref{fig:losses}. For $J/\Gamma=50K\: ns$ nearly
perfect transfer is still possible for chains up to a length of $N\approx40$.
In a single Heisenberg chain using the scheme described in \cite{SB03},
this system could only achieve a fidelity of $0.23$ when transferring
an exitation.%

Even if the amplitude damping is not symmetric, its effect is weaker
than in single spin schemes. This is because it can be split in a
symmetric and asymmetric part. The symmetric part can be overcome
with the above strategies. For example, if the amplitude damping on
the chains is $\Gamma_{1}$ and $\Gamma_{2}$ with $\Gamma_{1}>\Gamma_{2},$
the state (\ref{eq:jump1}) will be\begin{eqnarray}
 &  & \sum_{n=1}^{N}f_{n,1}(t)\left\{ \alpha e^{-\Gamma_{2}t}\left|\mathbf{0n}\right\rangle +\beta e^{-\Gamma_{1}t}\left|\mathbf{n0}\right\rangle \right\} \\
 & = & e^{-\Gamma_{2}t}\sum_{n=1}^{N}f_{n,1}(t)\left\{ \alpha\left|\mathbf{0n}\right\rangle +\beta e^{-\left(\Gamma_{1}-\Gamma_{2}\right)t}\left|\mathbf{n0}\right\rangle \right\} \\
 & \approx & e^{-\Gamma_{2}t}\sum_{n=1}^{N}f_{n,1}(t)\left|\mathbf{s}(n)\right\rangle \label{eq:nodeviation}\end{eqnarray}
provided that $t\ll\left(\Gamma_{1}-\Gamma_{2}\right)^{-1}.$ Using
a chain of length $N=20$ with $J=20K*k_{B}$ and $\Gamma_{1}^{-1}=4ns$,
$\Gamma_{2}^{-1}=4.2ns$ we would have to fulfill $t\ll164ns$. We
could perform approximately $10$ measurements (cf. Eq. (\ref{eq:timescale}))
without deviating too much from the state (\ref{eq:nodeviation}).
In this time, we can use our protocol in the normal way. The resulting
success probability given by the finite version of Eq. (\ref{eq:plimit})
would be $75$\%. A similar reasoning is valid for phase noise, where
the environment can be split into common and seperate parts. If the
chains are close, the common part will dominate and the seperate parts
can be neglected for short times.

Finally, let us mention another advantage of our scheme. In single
chain schemes, Bob has to extract the state precisely at an optimal
time to obtain it with high fidelity. Our scheme is robust to the
errors in this. Even if Bob measures to extract his state at an incorrect
(non-optimal) time, he will receive the perfect state conditional
on his measurement outcome. If he is unsuccessful, he simply tries
again, without having Alice to resend. Also, due to the conclusive
nature of the protocol, once Bob has received the state, the rest
of the channel is automatically in the ground state and does not need
to be reset for the next transfer (as opposed to many of the existing
schemes \cite{SB03,key-1,key-4,FVMAM04}).

\section{Conclusions\label{sec:Conclusions}}

In conclusion, we have presented a simple and efficient scheme for
conclusive and arbitrarily perfect quantum state transfer. To achieve
this, two parallel spin chains (individually amplitude damping channels)
have been used as one \emph{amplitude delaying channel}. We have shown
that our scheme is more robust to decoherence and imperfect timing
than the single chain schemes. Even though the encoding is simple,
it has made spin-chain based communications \emph{both} realistic
and perfect at the same time.

Our strategy can be generalized to graphs interconnecting many different
users, and to many other systems. As an example, we will now briefly
mention how our scheme can be adapted to a \emph{single} chain of
qutrits. The correct generalization of the exchange interaction for
a chain of 3-level quantum system is a SU(3) chain \cite{BS75}. For
example, a chain of atoms with three internal levels, $\left|-1\right\rangle $,
$\left|0\right\rangle $ and $\left|1\right\rangle $ in an optical
lattice, where the atoms can hop from site to site but more than one
atom cannot occupy a single site, will form a SU(3) chain. If we relabel
our parallel spin chain states $\left|\mathbf{0}\right\rangle ^{(1)}\left|\mathbf{0}\right\rangle ^{(2)}$
by $\left|0_{1}\ldots0_{n}\ldots0_{N}\right\rangle $, $\left|\mathbf{n}\right\rangle ^{(1)}\left|\mathbf{0}\right\rangle ^{(2)}$
by $\left|0_{1}\ldots1_{n}\ldots0_{N}\right\rangle $ and $\left|\mathbf{0}\right\rangle ^{(1)}\left|\mathbf{n}\right\rangle ^{(2)}$
by $\left|0_{1}\ldots-1_{n}\ldots0_{N}\right\rangle $, then our protocol
can be mapped to a single chain of qutrits interacting via SU(3) exchange.
Though the state is no longer encoded in a decoherence free subspace
as before, in an optical lattice implementation, one can use three
hyperfine ground states of atoms as $\left|-1\right\rangle $, $\left|0\right\rangle $
and $\left|1\right\rangle $ to completely avoid amplitude damping.

This work was supported by the UK Engineering and Physical Sciences
Research Council grant Nr. S627961/01 and the QIPIRC.

\end{document}